# Errors in Length-weight Parameters at FishBase.org

Simeon Cole-Fletcher, Lucas Marin-Salcedo, Ajaya Rana, and Michael Courtney
U.S. Air Force Academy, 2354 Fairchild Drive, USAF Academy, CO 80840
Michael.Courtney@usafa.edu

**Background:** FishBase.org is an on-line database of fish related data that has been cited over 1500 times in the fisheries literature. Length-weight relationships in fish traditionally employ the model, $W(L) = aL^b$, where L is length and W is weight. Parameters a and b are catalogued by FishBase for a large number of sources and species. FishBase.org detects outliers in a plot of log(a) vs. b to identify dubious length-weight parameters.
**Materials and Methods:** To investigate possible errors, length-weight parameters from FishBase.org were used to graph length-weight curves for six different species: channel catfish (*Ictalurus punctatus*), black crappie (*Pomoxis nigromacalatus*), largemouth bass (*Micropterus salmoides*), rainbow trout (*Oncorhynchus mykiss*), flathead catfish (*Pylodictis olivaris*), and lake trout (*Salvelinus namaycush*) along with the standard weight curves (Anderson and Neumann 1996, Bister et al. 2000). Parameters noted as "doubtful" by FishBase were excluded. For each species, variations in curves were noted, and the minimum and maximum predicted weights for a 30 cm long fish were compared with each other and with the standard weight for that length. For lake trout, additional comparisons were made between the parameters and study details reported in FishBase.org for 6 of 8 length-weight relationships and those reported in the reference (Carlander 1969) for those 6 relationships.
**Results:** In all species studied, minimum and maximum curves produced with the length-weight parameters at FishBase.org are notably different from each other, and in many cases predict weights that are clearly absurd. For example, one set of parameters predicts a 30 cm rainbow trout weighing 44 g. For 30 cm length, the range of weights (relative to the standard weight) for each species are: channel catfish (31.4% to 193.1%), black crappie (54.0% to 149.0%), largemouth bass (28.8% to 130.4%), rainbow trout (14.9% to 113.4%), flathead catfish (29.3% to 250.7%), and lake trout (44.0% to 152.7%). Ten of the twelve extreme curves reference two sources (Carlander 1969 and Carlander 1977). These two sources are used for a total of 100 different species at FishBase.org. In the case of lake trout, comparing the length-weight table at FishBase.org and the cited source (Carlander 1969) revealed that while 5 of 6 total length measurements were incorrectly reported as fork lengths by FishBase.org, all parameters accurately reflected the source. Comparing the length-weight relationships of the source (Carlander 1969) with the table of weights in different length ranges reveals the length-weight parameters in the source are clearly in error. However, FishBase.org also neglects to specify clearly distinguished subspecies and/or phenotypes such as siscowet and humper lake trout.
**Conclusion:** Length-weight tables at FishBase.org are not generally reliable and the on-line database contains dubious parameters. Assurance of quality probably will require a systematic review with more careful and comprehensive methods than those currently employed.

## I. Introduction

The traditional power law model, $W(L) = aL^b$, finds widespread application for length-weight relationships in fish. Many studies simply measure weight and length of a number of samples, take the logarithms of length and weight and estimate best-fit parameters by means of linear least-squares (LLS) regression. (Anderson and Neumann 1996) Alternatively, the parameters can be estimated by the Levenberg-Marquardt non-linear least-squares (NLLS) method. The exponent, b, is usually close to 3.0, (Froese 2006) is independent of the system of units, and has an easily interpreted physical meaning as related to isometric growth for b = 3. (Pauly 1984) In contrast, the coefficient, a, depends strongly on both the exponent and units, and its physical meaning is difficult to interpret. FishBase.org has been cited over 1500 times; one paper presenting a detailed analysis of length-weight parameters at FishBase.org has been cited over 100 times. (Froese 2006)

Length-weight parameter values are catalogued at FishBase.org for the traditional model where the length is in cm and weight in g. The purpose of the present study is to identify possible errors by investigating the variations in weight predicted by different parameters for six species: channel catfish (*Ictalurus punctatus*), black crappie (*Pomoxis nigromacalatus*), largemouth bass (*Micropterus salmoides*), rainbow trout (*Oncorhynchus mykiss*), flathead catfish (*Pylodictis olivaris*), and lake trout (*Salvelinus namaycush*).

### II. Method

To investigate reliability in weight predicted by a given length, length-weight parameters obtained from FishBase.org are used to graph a number of length-weight curves for each species. Parameters noted as "doubtful" by FishBase and parameters obtained by studies measuring standard length are excluded. Variations in curves are easily visible when several length-weight curves are plotted together.

Standard weight curves are available for the species considered in the present study. These curves estimate the weight of fish in the 75$^{th}$ percentile for a given total length. (Anderson and Neumann 1996, Bister et al. 2000) The weight of 30 cm long fish are compared for the minimum, maximum, and standard weight curves. In lake trout, the source cited by FishBase.org for six length-weight relationships (Carlander 1969) is compared with the information reported by FishBase.org. New (accurate) length-weight parameters are determined both by linear least-squares (LLS) regression and non-linear least-squares (NLLS) regression for one set of data (*Salvelinus namaycush siscowet*) reported in Carlander (1969). Models from FishBase.org, Carlander (1969) and the new length-weight models are compared.

### III. Results

**1. Channel Catfish (*Ictalurus punctatus*)**

Length-weight curves for channel catfish are shown in Figure 1. The top curve has parameters a = 0.0041 and b = 3.407. These parameters originated in an Oklahoma study with 4617 samples. Fishbase.org cites Carlander (1969) as its source. This length-weight relationship implies a channel catfish with a total length of 30 cm will weigh 444.1 g, which is 193.1% of the standard weight at this length. (Anderson and Neumann 1996) This is unlikely.

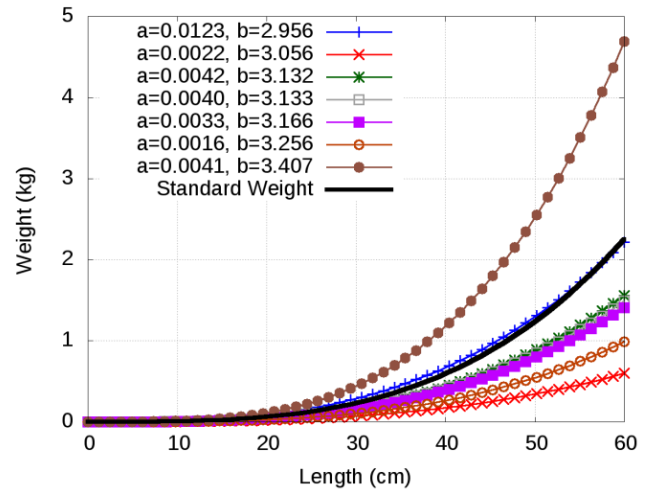

Figure 1: Weight vs. length curves from parameters at FishBase.org for channel catfish are compared with each other and with standard weight curve.(Anderson and Neumann 1996)

The bottom curve in Figure 1 has parameters a = 0.0022 and b = 3.056 obtained from an original study with 154 samples. The same source is cited. (Carlander 1969) Though the type of length measurement is not specified explicitly, this length-weight curve implies a channel catfish with a total length of 30 cm will weigh 70.23 g, which is 31.4% of the standard weight. Of course, the parameters seem even more errant if a fork length or standard length relationship is intended, because one would expect a fish with a standard or fork length of 30 cm to be even heavier than one 30 cm in total length.

**2. Black Crappie (*Pomoxis nigromacalatus*)**

Figure 2 shows length-weight curves for black crappie. The top curve has parameters a = 0.0195 and b = 3.081. Fishbase.org cites Carlander (1977) as its source. This curve suggests that a 30 cm fish will have a weight of 693.5 g which is 149.0% of the standard weight at this length.(Anderson and Neumann 1996) Since the type of length measurement is not specified, this curve is not completely absurd, and the weight values would be reasonable if the length were the standard length.

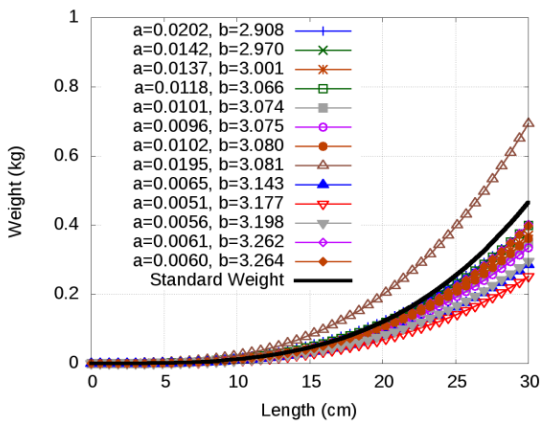

Figure 2: Select weight vs. length curves from parameters at FishBase.org for black crappie are compared with each other and with standard weight curve.(Anderson and Neumann 1996)

The bottom curve in Figure 2 has parameters a = 0.0051 and b = 3.177 and also cites Carlander (1977) as its source. Though the type of length measurement is not specified explicitly, this length-weight curve implies a black crappie with a total length of 30 cm will weigh 250.43 g, which is 54.0% of the standard weight.

### 3. Largemouth Bass (*Micropterus salmoides*)

Weight-length curves are shown for largemouth bass in Figure 3. The top curve has a = 0.0263 and b = 2.900. The original study was at Big Creek Reservoir, Iowa. FishBase.org cites Carlander (1977). The top curve predicts that a fish of 30 cm total length will weigh 505.37 g, which is 130.4% of the standard weight. This is unlikely, but not completely unreasonable in a population that is eating very well when the original study was done.

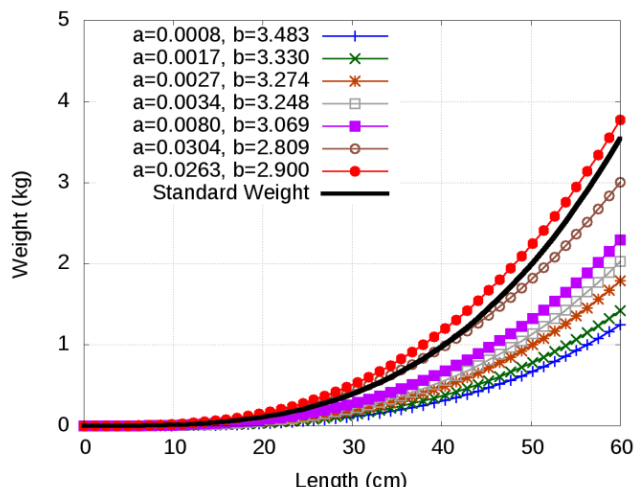

Figure 3: Select weight vs. length curves from parameters at FishBase.org for largemouth bass are compared with each other and with standard weight curve.(Anderson and Neumann 1996)

The bottom curve in Figure 3 has a = 0.0008 and b = 3.483. The original study was done in Bull Shoals Lake, Arizona in 1968. FishBase.org cites Carlander (1977). These parameter values predict a bass with a total length of 30 cm will weigh 110.27 g, which is 28.8% of the standard weight. A population of largemouth bass 30 cm long and 110 g are not skinny, poorly fed, or undernourished; they simply do not exist (until filleted). Length type is unspecified in this case; of course, the numbers are even more absurd if fork length or standard length is intended.

### 4. Rainbow Trout (*Oncorhynchus mykiss*)

Figure 4 shows length-weight curves for rainbow trout. The top curve has a coefficient a = 0.0089 and exponent b = 3.096. This study of 111 samples from Iran (Esmaeli and Ebrahimi 2006) predicts that a trout with 30 cm total length will weight 333.09 g, which is 113.4% of standard weight. This agrees with reasonable expectations and there is no reason to suspect error. In contrast, the bottom curve has coefficient a = 0.0063 and exponent b = 2.604. This study of an unknown number of samples and an unknown length type originated in Castle Lake, California. FishBase.org cites Carlander (1969). This curve suggests a typical 30 cm long trout will only weigh 43.9 g, which is 14.9% of standard weight.

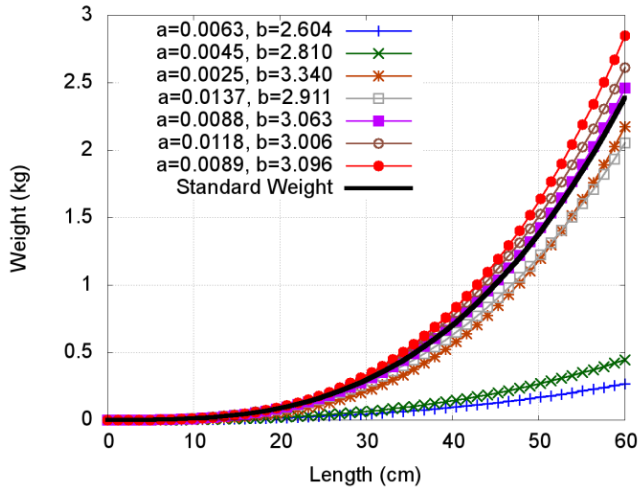

Figure 4: Select weight vs. length curves from parameters at FishBase.org for rainbow trout are compared with each other and with standard weight curve.(Anderson and Neumann 1996)

### 5. Flathead Catfish (*Pylodictis olivaris*)

Length-weight relationships for flathead catfish are shown in Figure 5. The top curve originates from a study in Grand Lake, Oklahoma and was produced with parameters a = 0.0121 and b = 3.233 obtained from FishBase.org, which cites Carlander (1969) as its source. This curve implies that a 30 cm fish will weigh 721.64 g, which is 250.7% of the standard weight, which seems unreasonably large even if the standard length is intended. (FishBase.org neither specifies the type of length nor the number of samples in this study.) The bottom curve is generated with paramaters a = 0.0007 and b = 3.440 obtained from FishBase.org, which also specifies total length and a sample size of 26 for a study conducted in Alabama. FishBase cites Carlander (1969) for this absurdity which implies a 30 cm long catfish might weigh 84.4 g, which is only 29.3% of its standard weight.

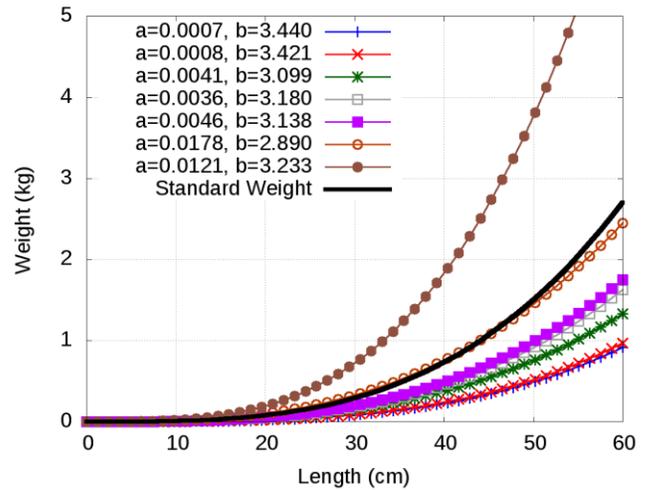

Figure 5: Weight vs. length curves from parameters at FishBase.org for flathead catfish are compared with each other and with standard weight curve.(Bister et al. 2000)

### 6. Lake Trout (*Salvelinus namaycush*)

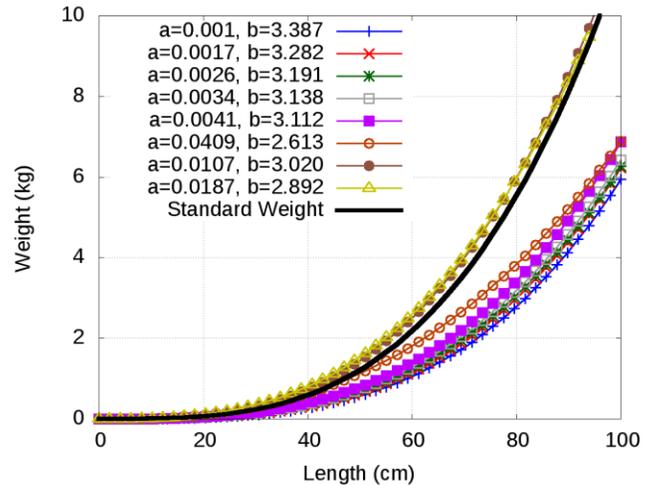

Figure 6: Weight vs. length curves from parameters at FishBase.org for lake trout are compared with each other and with standard weight curve.(Anderson and Neumann 1996)

Length-weight curves for lake trout are shown in Figure 6. The top curve has coefficient, a = 0.0187 and exponent b = 2.892. FishBase.org cites a study in Keller Lake, Canada (Johnson 1973) for these parameters which imply a trout with a 30 cm fork length will weight 349.68 g, which is 152.7% of the standard length if no adjustment is made from fork length to total length. Adjusting the standard weight curve with FL= TL/1.073 (Froese and Pauly 2010) suggests a relative weight of 124.6%, which is probably reasonable.

In contrast, the bottom curve is generated from parameters a = 0.001 and b = 3.387 obtained from FishBase.org which cites Carlander (1969). This Lake Superior study measured 393 specimens and FishBase.org specifies fork length as the type of length; however, Carlander (1969) reports total length. The parameter values at FishBase.org agree with Carlander (1969) for the significant digits reported, yet they imply that a lake trout of 30 cm length will weigh only 100.7 g, which is 44.0% the standard weight. (Anderson and Neumann 1996) It is also dubious that the only length-weight curve for the siscowet subspecies, which is known for fatness, would be the lowest available curve for length-weight in lake trout at FishBase.org. FishBase.org failed to note the subspecies, but it is clear in Carlander (1969).

Carlander (1969) also has a table of mean weights for 25 mm length groups. There was only one fish in the 279-305 mm length group, and its weight is reported as 272 g, which agrees with expectation that the siscowet phenotype is fatter than ordinary lake trout, but which disagrees with the implication of the length-weight parameters of 100.7 g weight for a 30 cm fish. Figure 7 shows the length-weight curve for the parameters reported by FishBase.org and Carlander (1969) with data from the mean weight vs. length range table from the same study, also reported in Carlander (1969). The errant parameter(s) clearly were present in Carlander (1969). Error bars in the figure are estimated as the extreme spread of weights divided by the square root of the number of samples. In cases where the extreme spread is unavailable or the sample size is 1, the assigned uncertainty is 10 percent of the mean weight divided by the square root of the number of samples. The real data are above the standard weight curve for lake trout (Anderson and Neumann 1996), as expected for siscowet lake trout.

Since the length-weight parameters reported by Carlander (1969) and FishBase.org are in error, accurate parameters have been determined both by non-linear least squares (NLLS) and linear-least squares (LLS) regression of the traditional length-weight model in fish, as well as for an improved model, $W(L) = (L/L_1)^b$, where b is the exponent, and $L_1$ is a new parameter representing the typical length of a fish that weighs 1 kg. Table 1 reports best fit parameters, error estimates, parameter covariances, and correlation coefficients for the models and regression techniques.

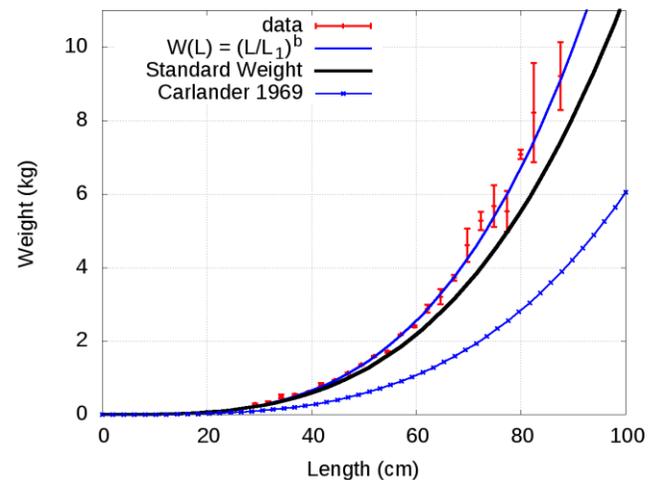

*Figure 7: Weight vs. length curves from parameters at FishBase.org agree with the length-weight relationship for the siscowet subspecies/phenotype of lake trout (lowest curve) published by Carlander (1969), but not with the data summarized by Carlander (1969) for the mean weight for each 25 mm increment. The new best-fit models agree with the data. Both data and new models are above the standard weight curve for lake trout (Anderson and Neumann 1996), as expected for a variety of lake trout noted for being fatter than ordinary lake trout. The mean weight for the 635-659 mm length range (2717 g) was obviously in error, as it was outside the range of weights (2853 g to 3556 g). The value reported by Carlander (1969) was replaced with the midpoint of the minimum and maximum weights in the 635-659 mm length range.*

| NLLS improved $W(L)=(L/L_1)^b$ | $L_1$ (cm) | 45.5164 |
|---|---|---|
| | $L_1$ error | 0.26% |
| | b | 3.3755 |
| | b error | 1.22% |
| | covariance | 0.6760 |
| | r | 0.9965 |
| NLLS traditional $W(L)=aL^b$ | **a** | **0.00253** |
| | a error | 16.26% |
| | b | 3.3755 |
| | b error | 1.21% |
| | covariance | -0.9990 |
| | r | 0.9965 |
| LLS traditional $W(L)=aL^b$ | **a** | **0.00385** |
| | a error | 18.71% |
| | b | 3.2796 |
| | b error | 1.42% |
| | covariance | -0.9970 |
| | r | 0.9963 |

*Table 1: Results of non-linear least squares (NLLS) fitting of the siscowet lake trout data from Carlander (1969) to both the traditional length-weight model and an improved model. The improved model uses a parameter, $L_1$, which is the typical length of a fish weighing 1 kg. The improved model has about the same correlation coefficient and estimated uncertainty in the parameter b compared with the traditional model, but the uncertainty in $L_1$ is much smaller (0.26%) than the estimated uncertainties in the parameter a using either the NLLS regression of linear least-squares (LLS) regression of log(W) vs. log(L). The covariance between parameters in the improved model is also smaller in magnitude than the covariance between parameters in the traditional model.*

## IV. Discussion

FishBase.org attempts to detect errors in length-weight parameters by plotting log(a) vs. b and looking for outliers from the line formed for parameters available for a given species. (Froese and Pauly 2010) Figure 18 of Froese and Pauly (2010) shows an example plot for largemouth bass. However, the application of this procedure has failed to identify likely errors in parameters apparently originating in Carlander (1969) and Carlander (1977), as well as failing to identify cases of fork length being substituted for total length. FishBase.org reports that Carlander (1969) is a source for 84 species and Carlander (1977) is a source for an additional 16.

Since some of the reported weight-length parameters predict weights far from reality, it seems plausible that at least some of the errors arise not from misidentification of fish species, small sample sizes, or fish populations deviating significantly from typical. Transcription errors and errors in unit conversions in reporting from the original source seem more likely in these cases. Regardless of the source of these errors, their prevalence suggests more reliable error detection is needed before listing length-weight parameters at FishBase.org, and before using bulk length-weight parameters for additional studies.(Froese 2006) Plotting of weight vs. length curves and comparing predicted weights for 30 cm long fish with reliable measurements or weights predicted by standard weight curves should help to recognize obvious outliers. Identifying more subtle errors probably requires repeating regression analyses either with the original data or with data for average weights in a given length interval.

An improved model is suggested which introduces a parameter with clear physical meaning (the typical length of a fish weighing 1 kg). Egregious errors in a more meaningful parameter should be less likely to go unnoticed for many years. This improved model yields a smaller magnitude covariance with the exponent as well as a much smaller estimate of its uncertainty than the coefficient, a. All the length-weight parameters reported by Carlander (1969) and Carlander (1977) should be reviewed for errors, preferably by comparing with the source data and/or repeating the regression analysis, where possible. The length-weight parameters reported by FishBase.org should be reviewed by a more reliable method than currently employed. The impact of existing errors on subsequent work should be carefully considered in each case.

Finally, these errors bring to mind the question, "Who is responsible for errors in scientific reporting?" When an error in an original source is repeated by subsequent sources citing the original, does the sole responsibility for all subsequent citations rest with the original source, or should subsequent authors and editors bear some responsibility for due diligence in error detection prior to repeating previously published information?

## V. Acknowledgements

The authors acknowledge helpful discussions with Amy Courtney, PhD (BTG Research) and Beth Schaubroeck, PhD (USAFA Department of Mathematical Sciences). We thank the Quantitative Reasoning Center at the United States Air Force Academy for supporting this work. We are also grateful to three anonymous peer reviewers who offered helpful suggestions for improving this paper.

## VI. References

**Anderson R.O., Neumann R.M.** 1996. Length, Weight, and Associated Structural Indices, Pp. 447-481. *In:* Murphy B.E. and Willis D.W. (eds.) Fisheries Techniques, second edition. American Fisheries Society.

**Bister T.J., Willis D.W., Brown M.L., Jordan S.M., Neumann R.M., Quist M.C., Guy C.S.** 2000. Proposed Standard Weight ($W_s$) Equations and Standard Length Categories for 18 Warmwater Nongame and Riverine Fish Species. North American Journal of Fisheries Management **20**:570–574.

**Carlander K.D.** 1969. Handbook of freshwater fishery biology, volume 1., The Iowa State University Press, Ames. Iowa.

**Carlander K.D.** 1977. Handbook of freshwater fishery biology, volume 2. The Iowa State University Press, Ames, Iowa.

**Froese R.** 2006. Cube law, condition factor and weight–length relationships: history, meta-analysis and recommendations. Journal of Applied Ichthyology, 22: 241–253. doi: 10.1111/j.1439-0426.2006.00805.x

**Froese R., Pauly D.** Editors. 2010. FishBase. World Wide Web electronic publication. www.fishbase.org, version (09/2010).

**Esmaeli H.R., M. Ebrahimi** 2006. Length-weight relationships of some freshwater fishes of Iran. J. Appl. Ichthyol. 22:328-329.

**Johnson L.** 1973. Stock and recruitment in some unexploited Canadian Arctic lakes. Rapp. P.-v. Réun. Cons. int. Explor. Mer 164:219-227.

**Pauly D.** 1984. Fish Population Dynamics in Tropical Waters. International Center for Living Aquatic Resources Management. Manila.